\documentclass[a4paper,11pt]{article}
\usepackage{pos}
\usepackage{amsmath,bm,color,fancyhdr,hyperref,lastpage,units,soul,caption}
\usepackage{graphicx}
\graphicspath{figures/}
\title{Charm Fluctuations and Deconfinement}
\author*[a,1]{Sipaz Sharma}
\affiliation[a]{Fakult\"at f\"ur Physik, Universit\"at Bielefeld, D-33615 Bielefeld, Germany}
\note{For the HotQCD Collaboration.}

\abstract{We establish that the charmed hadrons start dissociating at the chiral crossover temperature, ${T_{pc}}$, leading to the appearance of charm degrees freedom carrying fractional baryon number. Our method is based on analyzing the second and fourth-order cumulants of charm (${C}$) fluctuations, and their correlations with baryon number (${B}$), electric charge (${Q}$) and strangeness (${S}$) fluctuations. The first-time calculation of the ${QC}$ correlations on the high statistics datasets of the HotQCD Collaboration enables us to disentangle the contributions from different electrically-charged charm subsectors in the hadronic phase. In particular, we see an enhancement over the PDG expectation in the fractional contribution of the ${|Q|}=2$ charm subsector to the total charm partial pressure for ${T<T_{pc}}$; this enhancement is in agreement with the Quark Model extended Hadron Resonance Gas (QM-HRG) model calculations. Furthermore, the agreement of QM-HRG calculations with the projections onto charmed baryonic and mesonic correlations in different charm subsectors indicates the existence of not-yet-discovered charmed hadrons in all charm subsectors below ${T_{pc}}$. We aim at determining the relevant degrees of freedom in temperature range ${T_{pc}<T<340 \text{ MeV}}$ by assuming the existence of a non-interacting gas of charmed quasi-particles composed of meson, baryon and quark-like excitations above $T_{pc}$. Our data suggest that the particles with quantum numbers consistent with quarks start appearing at $T_{pc}$. }
	
\FullConference{The 40th International Symposium on Lattice Field Theory (Lattice 2023)\\
July 31st - August 4th, 2023\\
Fermi National Accelerator Laboratory\\}

\begin{document}
\maketitle

\section{Introduction}

The fact that charmed hadrons start melting at the chiral crossover temperature, ${T_{pc}=156.5\pm1.5}$ MeV \cite{HotQCD:2018pds, PhysRevLett.125.052001,Bazavov:2023xzm}, has been confirmed over the years by comparison of lattice QCD calculations with Hadron Resonance Gas (HRG) model predictions \cite{BAZAVOV2014210, Sharma:2022ztl}. 
 However, the nature of relevant charmed degrees of freedom above $T_{pc}$ remains elusive. For this purpose, we exploit the fact that, unlike light quarks, heavier quarks can be treated in the Boltzmann approximation. We construct various charm cumulants by calculating up to fourth-order fluctuations of charm ($C$), and their correlations with baryon number ($B$), electric charge ($Q$), and strangeness ($S$) fluctuations to understand the relevant degrees of freedom in the temperature range of interest. 
 
 Our main conclusions are based on the observation that, unlike absolute cumulants, the ratio of two cumulants is independent of the lattice spacing. However, the extraction of total charm pressure, $P_C$, and various partial charm pressures is crucial to understand the thermodynamics of strongly interacting matter. Therefore, we go one step further to understand the lattice cutoff effects in the absolute cumulants of the charm sector. We aim to approach the continuum limit by conducting independent calculations corresponding to two different Lines of Constant Physics (LCPs).
 
 \section{Boltzmann Approximation}
 \subsection{Charmed Hadrons}
Hadron resonance gas model (HRG) has been successful in describing the particle abundance ratios measured in the heavy-ion experiments. It describes a non-interacting gas of hadron resonances, and therefore can be used to calculate the hadronic pressure below ${T_{pc}}$ \cite{ANDRONIC2019759}. In the Boltzmann approximation, the dimensionless partial pressures from the charmed-meson, ${M_C},$  and the charmed-baryon, ${B_C}$, sectors
% related to total open-charm pressure, $P_C$, by
% \begin{equation}
%	{P_C(T,\overrightarrow{\mu})/T^4=M_C(T,\overrightarrow{\mu})+B_C(T,\overrightarrow{\mu})} \text{ ,}
%\end{equation}
take the following forms \cite{Allton:2005gk}:
\begin{gather}
	\begin{aligned}
		{M_C(T,\overrightarrow{\mu})}&{=\dfrac{1}{2\pi^2}\sum_{i\in \text{C-mesons}}g_i \bigg(\dfrac{m_i}{T}\bigg)^2K_2(m_i/T)\text{cosh}(Q_i\hat{\mu}_Q+S_i\hat{\mu}_S+C_i\hat{\mu}_C)} \text{ ,}\\
		{B_C(T,\overrightarrow{\mu})}&={\dfrac{1}{2\pi^2}\sum_{i\in \text{C-baryons}}g_i \bigg(\dfrac{m_i}{T}\bigg)^2K_2(m_i/T)\text{cosh}(B_i\hat{\mu}_B+Q_i\hat{\mu}_Q+S_i\hat{\mu}_S+C_i\hat{\mu}_C)} \text{ .}
		\label{eq:McBc}
	\end{aligned}
\end{gather}

In above equations, at a temperature ${T}$, the summation is over all charmed mesons/baryons (C-mesons/baryons ) with masses given by ${m_i}$; degeneracy factors of the states with equal mass and same quantum numbers are represented by ${g_i}$; to work with a dimensionless notation, chemical potentials corresponding to conserved quantum numbers are normalised by the temperature: ${\hat{\mu}_X = \mu/T}$, ${\forall X \in \{B, Q, S, C\}}$; ${K_2(x)}$ is a modified Bessel function, which for a large argument can be approximated by
${K_2(x)}\sim\sqrt{\pi/2x}\;e^{-x}\;[1+\mathbb{O}(x^{-1})]$\cite{Allton:2005gk}: consequently, if a charmed state under consideration is much heavier than the relevant temperature scale, such that ${m_i\gg T}$, then the contribution to ${P_C}$ from that particular state will be exponentially suppressed, e.g., the singly-charmed
${\Lambda}_c^{+}$ baryon has a Particle Data Group (PDG) mass of about $2286$ MeV, whereas the doubly-charmed ${\Xi_{cc}^{++}}$ baryon's mass as tabulated in PDG records is about $3621$ MeV, therefore at ${T_{pc}}$, the contribution to ${B_C}$ from ${\Xi_{cc}^{++}}$ will be suppressed by a factor of $10^{-4}$ in relation to ${\Lambda}_c^{+}$ contribution.

\subsection{Charm Quarks}
Charm quarks offer an advantage over the light quarks because for temperatures a few times $T_{pc}$, the Boltzmann approximation works for an ideal massive quark-antiquark gas. Therefore, in this approximation, the dimensionless partial charm quark pressure, $Q_C$ is given by, 
\begin{equation}
Q_C(T,\overrightarrow{\mu})=\dfrac{3}{\pi^2}\bigg(\dfrac{m_c}{T}\bigg)^2K_2(m_c/T)\text{cosh}\bigg(\dfrac{2}{3}\hat{\mu}_Q+\dfrac{1}{3}\hat{\mu}_B+\hat{\mu}_C\bigg) \text{ ,}
\end{equation}
where $m_c$ is the pole mass of the charm quark, and the degeneracy factor is 6.

\section{Generalized Susceptibilities of the Conserved Charges}
\label{sec: susc}
To project out the relevant degrees of freedom in the charm sector, one calculates the generalized susceptibilities, ${\chi^{BQSC}_{klmn}}$, of the conserved charges. This involves taking appropriate derivatives of the total pressure ${P}$, which contains contributions from the total charm pressure, $P_C(T,\vec{\mu})=\chi_2^C=\chi_n^C$, for $n$ even. These derivatives are taken with respect to the chemical potentials of the quantum number combinations one is interested in:
\begin{equation} 
	{\chi^{BQSC}_{klmn}=\dfrac{\partial^{(k+l+m+n)}\;\;[P\;(\hat{\mu}_B,\hat{\mu}_Q,\hat{\mu}_S,\hat{\mu}_C)\;/T^4]}{\partial\hat{\mu}^{k}_B\;\;\partial\hat{\mu}^{l}_Q\;\;\partial\hat{\mu}^{m}_S\;\;\partial\hat{\mu}^{n}_C}}\bigg|_{\overrightarrow{\mu}=0}\text{.}
	\label{eq:chi}
\end{equation}
To make R.H.S dimensionless, $P$ is normalized by $T^4$. If ${n \neq 0}$, ${P}$ can be replaced by ${P_C}$, since the derivative w.r.t. ${\hat{\mu}_C}$ will always project onto the open-charm sector. Note that ${\chi^{BQSC}_{klmn}}$ will be non-zero only for ${(k+l+m+n) \in \text{even}}$. In the following, if the subscript corresponding to a conserved charge is zero in the L.H.S. of Eq. \ref{eq:chi}, then both the corresponding superscript as well the zero subscript will be suppressed. Also, the terms cumulants, fluctuations and generalized susceptibilities will be used interchangebly throughout the text.

It is possible to construct the ratios of charm fluctuations defined in Eq. \ref{eq:chi}, which independent of the details of the hadron spectrum, take a particular value in the low temperature phase i.e., below $T_{pc}$, and attain a different value when states with baryon number other than $1$ or $0$ start appearing. Moreover, some ratios of charm fluctuations calculated in the framework of lattice QCD can receive enhanced contributions due the existence of not-yet-discovered open-charm states. It is possible to compare this enhancement to the HRG calculations performed with two data sets. The first scenario, denoted by PDG-HRG, is based on the states tabulated in the PDG records. The second scenario, denoted by QM-HRG, in addition to PDG states, takes into account states predicted via Quark-Model calculations \cite{Ebert:2009ua, Ebert:2011kk, Chen:2022asf}. Fig. \ref{fig:LCPs}~[right] shows the ratio of $\chi^{BC}_{13}$ and the total charm pressure. The former can be interpreted as $B_C$ in the validity range of HRG.  In Fig. \ref{fig:LCPs}~[right], the calculation based on the PDG-HRG states clearly misses the lattice data, whereas the QM-HRG dataset shows agreement with the lattice results, thus corroborating our previous findings on the not-yet-discovered charmed hadronic states -- with the majority of missing resonances being baryonic in nature \cite{BAZAVOV2014210, Sharma:2022ztl}.
\newpage
\section{Tuning the Charm Quark Mass on Lattice}

\label{sec:mass_tunning}
\subsection{Different Lines of Constant Physics}
\begin{minipage}[b]{0.45\textwidth}
	%\begin{figure}
	
	\includegraphics[width=\textwidth]{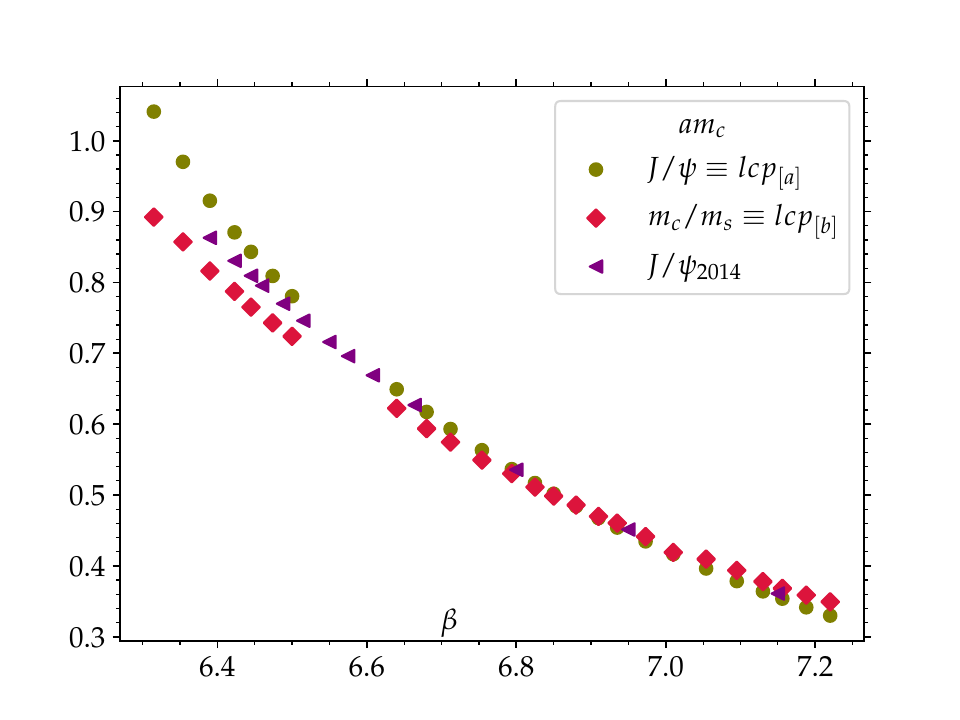}
	\captionof{figure}{Shown are the bare charm quark mass values as a function of the inverse gauge coupling, $\beta$, tuned using three different lines of constant physics: LCP[a], LCP[b] (see text), and the LCP presciption used in Ref.~\cite{BAZAVOV2014210}.}
	\label{fig:para_comp}
	%\end{figure}
\end{minipage}
\hspace{2 mm}
\begin{minipage}[b]{0.45\textwidth}
	We used two different Lines of Constant Physics (LCPs) to tune the bare charm quark mass $am_c$, where $a$ is the lattice spacing. The first LCP corresponds to keeping the spin-averaged charmonium mass, ${(3m_{J/\psi}+m_{\eta_{c\bar{c}}})/4}$, fixed to its physical value. Further details of the charm-quark mass tuning and parametrization can be found in \cite {Sharma:2022ztl}. The second LCP  is defined by the physical (PDG) charm to strange quark mass ratio, $m_c/m_s=11.76$ \cite{ParticleDataGroup:2022pth}.
	Results based on the above two LCPs will henceforth contain subscripts [a] and [b], respectively.  Results without any of these subscripts will correspond to LCP[b].
	\end{minipage}
	
\subsection{Cutoff Independence of the Ratios from Different LCPs}
\label{subsec:lcps}
 \begin{figure}[!h]
	\includegraphics[width=0.32\textwidth]{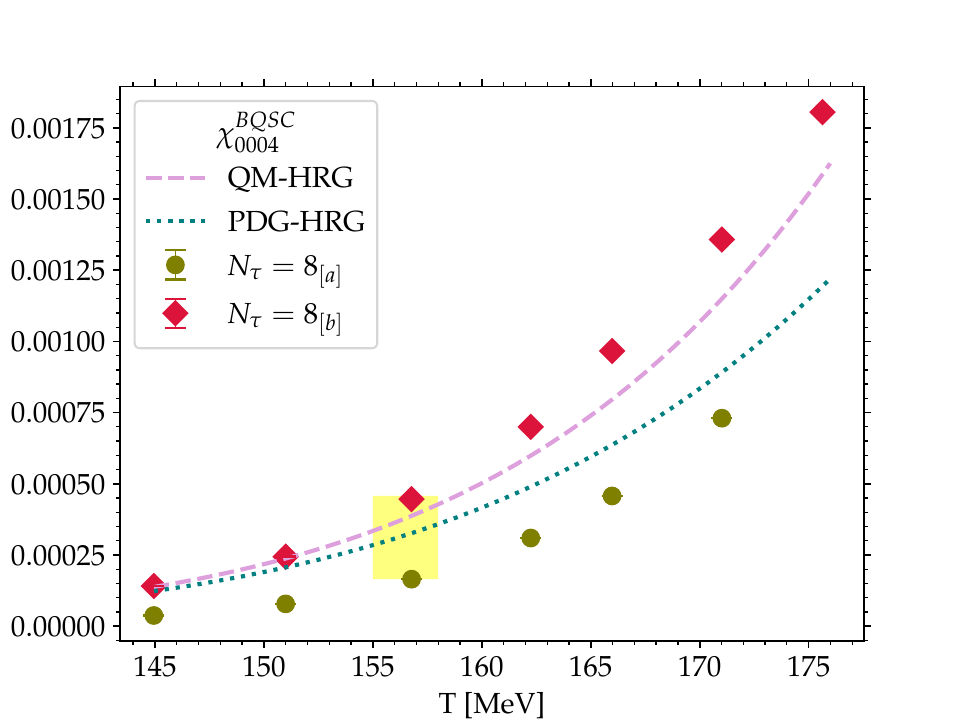}
	\hspace{0.01\textwidth}
	\includegraphics[width=0.32\textwidth]{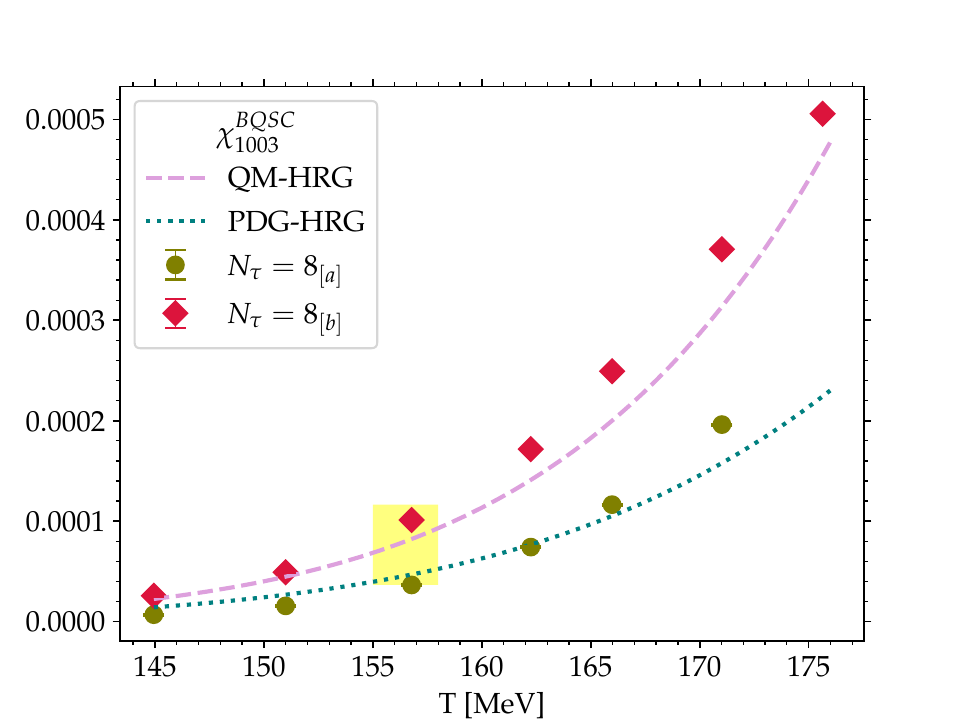}
	\hspace{0.01\textwidth}
	\includegraphics[width=0.32\textwidth]{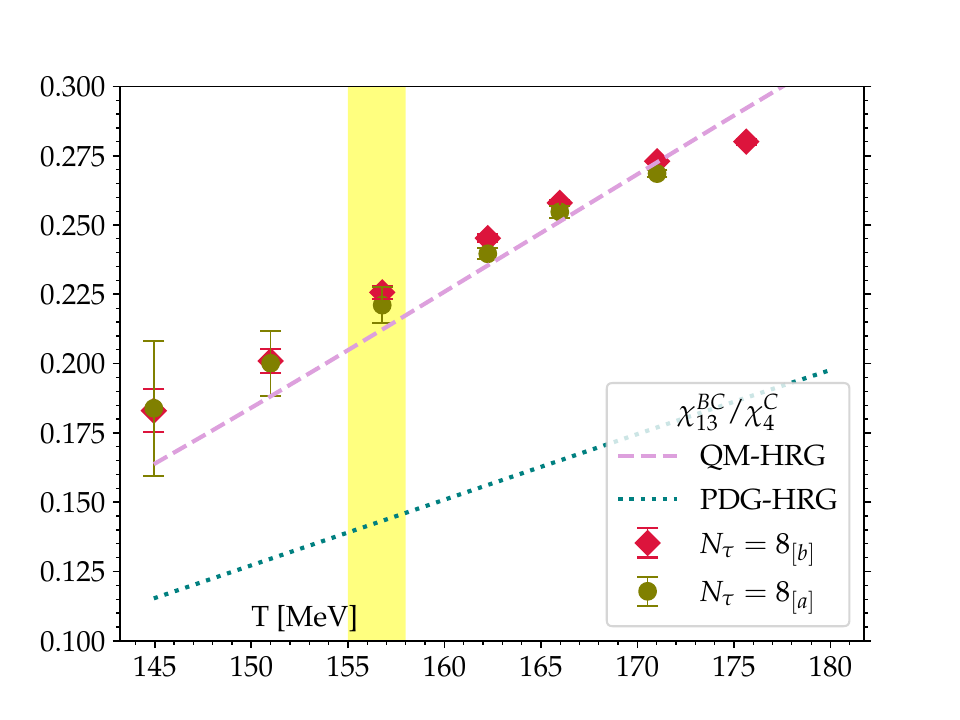}
	\caption{The total charm pressure as a function of temperature using two different LCP prescriptions at $N_\tau=8$ [left]. $\chi^{BC}_{13}$ as a function of temperature using two different LCP prescriptions at $N_\tau=8$  [middle]. The ratio $\chi_{13}^{BC}/\chi_4^C$ as a function of temperature obtained for LCP[a] and LCP[b] [right]. Also shown are the respective results obtained in PDG-HRG and QM-HRG model calculations. The yellow bands represent $T_{pc}$ with its uncertainty.}
	\label{fig:LCPs}
\end{figure}

The reason we chose to work with two different LCPs is that the absolute predictions in the charm sector are particularly sensitive to the precise tuning of the input bare quark masses. At a finite lattice spacing, different lattice cut-off effects for different hadrons lead to disagreement between absolute predictions of the two LCPs in Fig.~\ref{fig:LCPs}~[left] and~[middle]. In contrast to the absolute predictions, sensitivity to the choice of LCP cancels to a large extent in the ratios of the absolute predictions, and Fig.~\ref{fig:LCPs}~[right] is one such example. However, in addition to the ratios, we expect absolute predictions from LCP [a] and [b] to converge to the same values in the continuum limit.

\section{Appearance of Quark-Like Excitations near $T_{pc}$}
 \begin{figure}[!h]
	\includegraphics[width=0.32\textwidth]{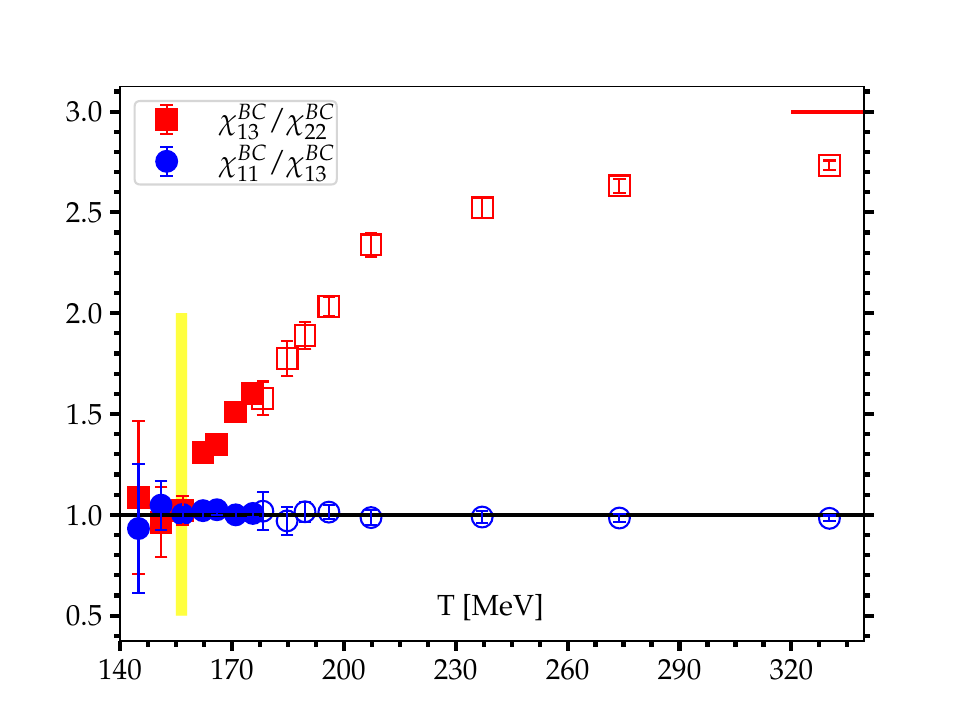}
	\hspace{0.01\textwidth}
	\includegraphics[width=0.32\textwidth]{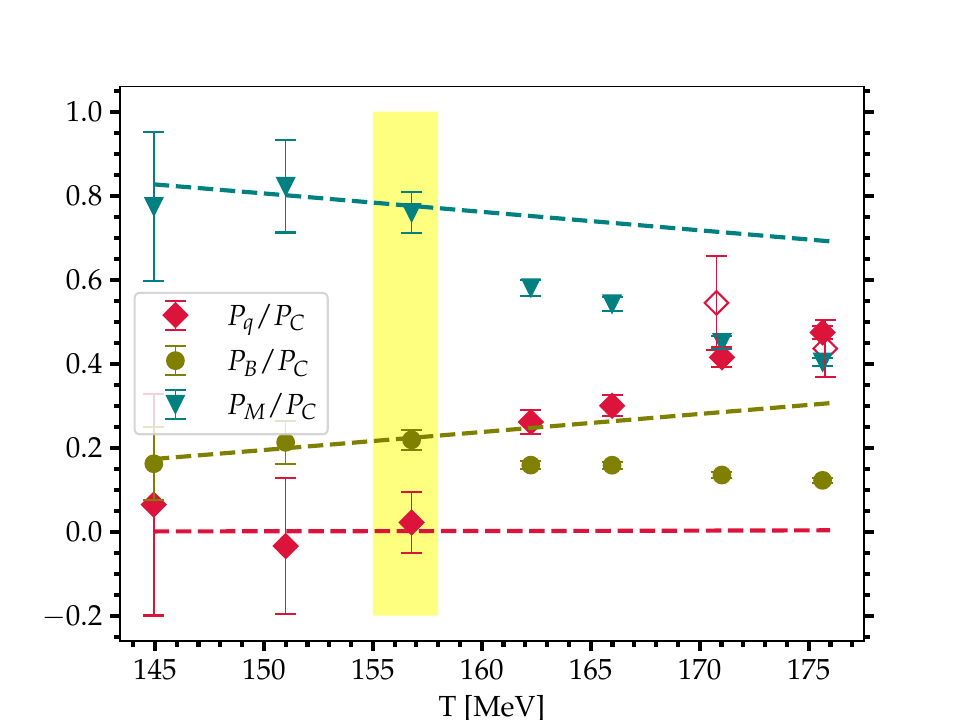}
	\hspace{0.01\textwidth}
	\includegraphics[width=0.32\textwidth]{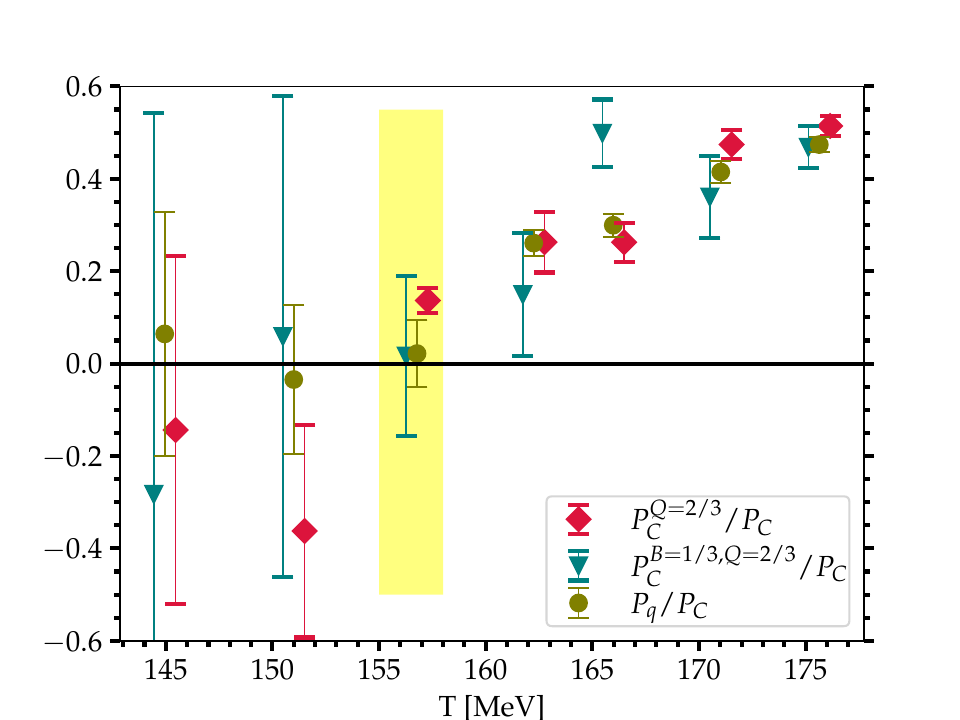}
	\caption{The ratios of different baryon-charm fluctuations as functions of temperature. The open symbols
		represent the results from Ref. \cite{BAZAVOV2014210}.
		The red solid line is the ideal charm quark gas limit of the ratio $\chi_{13}^{BC}/\chi_{22}^{BC}$ [left]. Partial pressures of charmed mesons, charmed baryons and charm quarks normalized by $P_C$ as functions of temperature. The dashed lines show corresponding results obtained from the QM-HRG model. The open symbols show the results for $N_{\tau}=12$ lattices [middle]. The partial pressures 
		of quasi-particles carrying
		(i) baryon number $|B|=1/3$ ($P_q$), (ii) electric charge
		$|Q|=2/3$ ($P_C^{Q=2/3}$), and
		(iii) ($|B|=1/3,\ |Q|=2/3$)  ($P_C^{B=1/3,Q=2/3}$) normalized by $P_C$.
		 Note that the T-coordinates of case (ii) and case (iii) are shifted by $\pm 0.51$ MeV respectively [right]. The yellow bands represent $T_{pc}$ with its uncertainty.}
	\label{fig:quarks}
\end{figure}
The ratios of $BC$ correlations with different numbers of $\hat{\mu}_B$ derivatives, such as $\chi^{BC}_{13}/\chi^{BC}_{22}$ shown in Fig.~\ref{fig:quarks}~[left], give a clear indication of hadron melting by deviating from unity. Slightly above $T_{pc}$, $\chi^{BC}_{13}/\chi^{BC}_{22}$ becomes greater than $1$, signalling the appearance of charm degrees of freedom carrying fractional $B$. For $T>300 \text{MeV}$, $\chi^{BC}_{13}/\chi^{BC}_{22}$ approaches its non-interacting quark-gas limit, which is $3$ -- shown with the solid-red line. Due to the dominance of the singly-charmed sector, $\chi^{BC}_{11}/\chi^{BC}_{13}$ shown in Fig.~\ref{fig:quarks}~[left] stays unity for the entire temperature range.

To understand the nature of charm degrees of freedom above $T_{pc}$, we extend the simple hadron gas model allowing the presence of partial charm quark pressure based on Ref.~\cite{Mukherjee:2015mxc}:
\begin{align}
	P_C(T,\vec{\mu})=M_C(T,\vec{\mu})+B_C(T,\vec{\mu})+Q_C(T,\vec{\mu}) \, .
	\label{eq:Pmodel}
\end{align}
For details please see our recent work \cite{Bazavov:2023xzm}. By considering only two quantum numbers: $B$ and $C$, the partial pressures of quark, baryon and meson-like excitations for $\mu=0$ can be expressed in terms of the generalized susceptibilties as follows,

\begin{align}
	P_{q}&=9(\chi^{BC}_{13}-\chi^{BC}_{22})/2\; , 
	\label{eq:partial-quasi} 	\\
	P_{B}&=(3\chi^{BC}_{22}-\chi^{BC}_{13})/2\; , 
	\label{eq:partial-quasiB}\\
	P_{M}&=\chi^{C}_{4}+3\chi^{BC}_{22}-4\chi^{BC}_{13} \; .
\end{align}
Upon breakdown of the HRG description at $T_{pc}$ in Fig. \ref{fig:quarks}~[middle], the fractional contribution of both the charmed mesonic and the charmed baryonic states to the total charm pressure starts decreasing in comparsion to their respective QM-HRG expectations, whereas the fractional contribution of the charmed states with $|B|=1/3$ becomes non-zero slighly above $T_{pc}$, and continues to increase as a function of temperature. To rule out the role of lattice cutoff effects in making $P_q/P_C$ non-zero, we also show $N_\tau=12$ results for the highest two temperatures using unfilled-red markers in Fig. \ref{fig:quarks}~[middle]. The $N_\tau=12$  results clearly agree with the $N_\tau=8$ results within errors, which further supports the presence of charm quark-like excitations in the Quark-Gluon Plasma (QGP).

To strengthen our claim, in addition to the partial quark pressure construction using $B$ and $C$ quantum numbers in Eq. \ref{eq:partial-quasi}, we independently construct partial pressure of quark-like excitations in two other ways. Firstly, by considering two quantum numbers: $Q$ and $C$. Secondly, by considering three quantum numbers: $B$, $Q$ and $C$. In addition to $P_q/P_C$, both these partial pressures normalised by $P_C$ are shown in Fig. \ref{fig:quarks}~[right], and they take the following forms:
\begin{align}
	P_{C}^{Q=2/3}&=\frac{1}{8}\big[54\chi^{QC}_{13}-81\chi^{QC}_{22}+27\chi^{QC}_{31}\big]\; ,\\
	P_{C}^{B=1/3,Q=2/3}&=\frac{27}{4}\big[\chi^{BQC}_{112}-\chi^{BQC}_{211}]\; .
\end{align}
Above partial pressures are constructed by assuming that only states carrying $|B|=0,1\text{ and }1/3$ contribute to $P_C$. This assumption is justified because our lattice data within errors satisfies the condition, $\chi_{13}^{BC} - 4\chi_{22}^{BC} + 3\chi_{31}^{BC} = 0$, see Ref.~\cite{Bazavov:2023xzm}. This implies four possibilities in the $QC$ sector: $|Q| = 0, 1, 2 \text{ and }2/3$, and three possibilities in the $BQC$ sector:  i) $|B|=1, |Q|=1$;  ii) $|B|=1, |Q|=2$;  iii) $|B|=1/3, |Q|=2/3$. Fig. \ref{fig:quarks}~[right] shows a remarkable agreement between three independent partial pressure constructions of quark-like excitations providing support to the quasi-particle model in Eq. \ref{eq:Pmodel}. Notice that $P_{C}^{Q=2/3}$ is senstive to contributions from $|Q|\neq 0, 1 \text{ and } 2$ charm sectors. The fact that it agrees with $P_q$ and $P_{C}^{B=1/3,Q=2/3}$ implies that these three quantities project onto the same quasi-particle sector.
\section{ $QC$ Sector below $T_{pc}$}
\begin{figure}[!h]
	\includegraphics[width=0.32\textwidth]{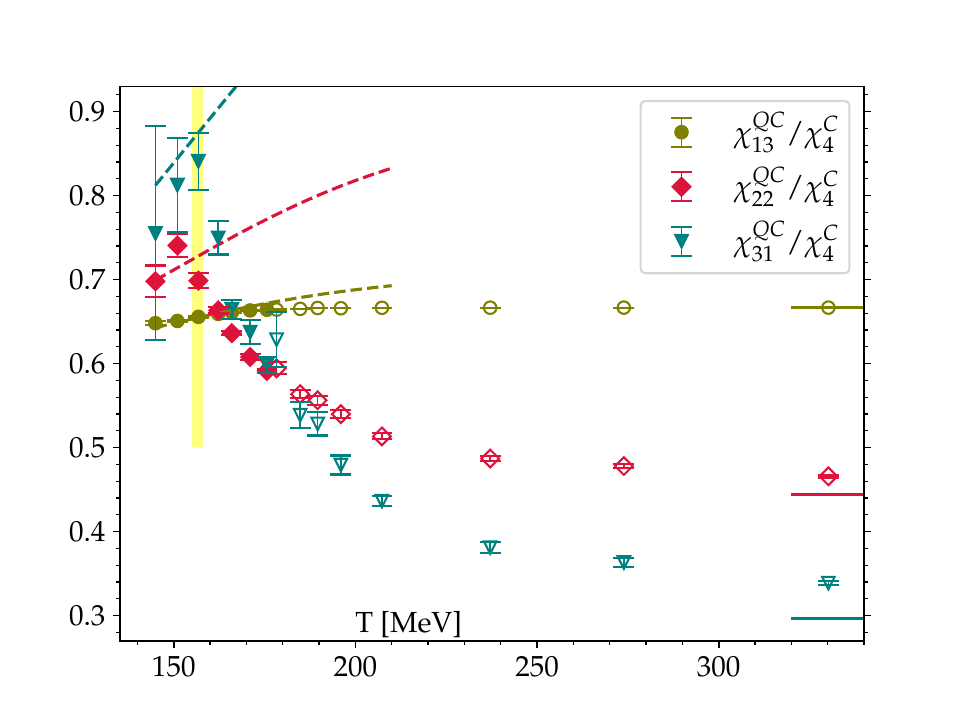}
	\hspace{0.01\textwidth}
	\includegraphics[width=0.32\textwidth]{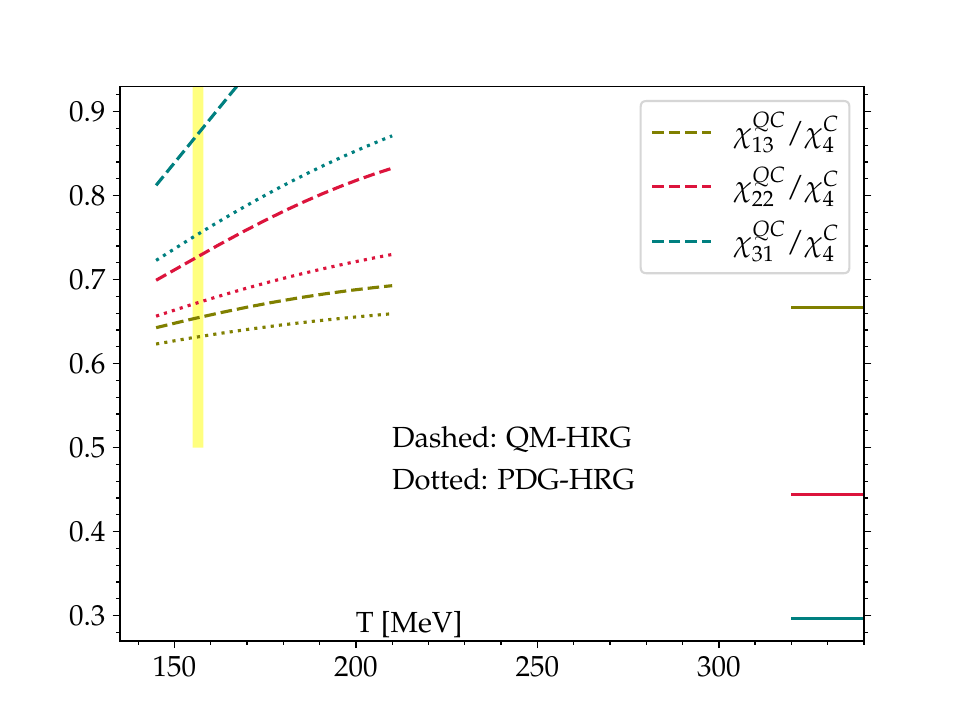}
	\hspace{0.01\textwidth}
	\includegraphics[width=0.32\textwidth]{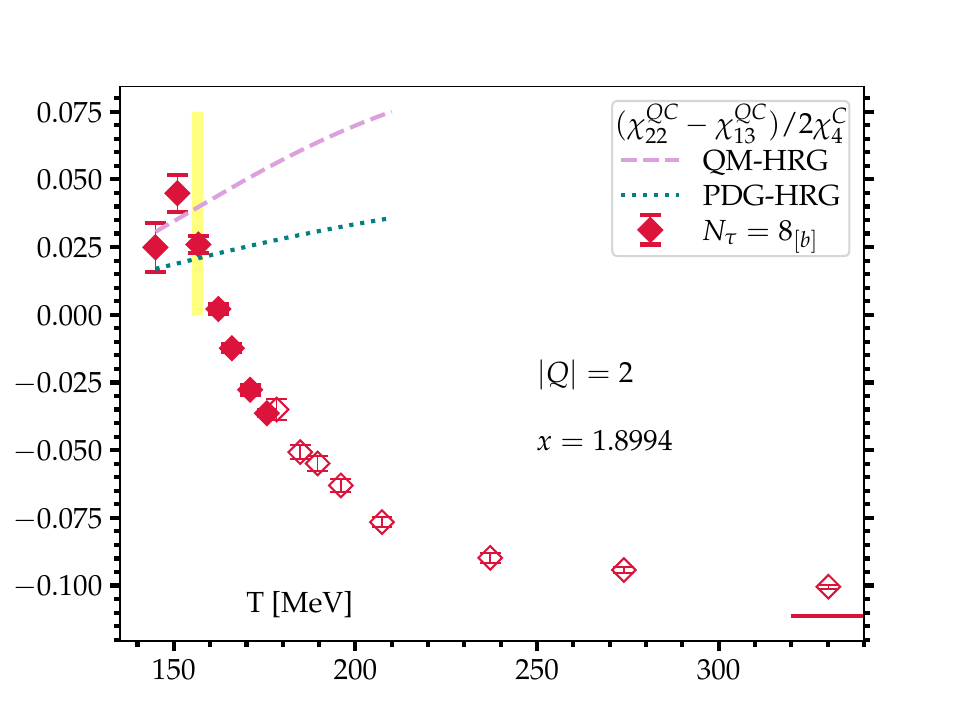}
	\caption{ The lattice results for the ratios of the fourth-order $QC$ correlations and $P_C$ as functions of temperature. The corresponding QM-HRG predictions are shown with the dashed lines of the same colors, whereas the solid lines represent the respective ideal charm quark gas limits [left]. The QM-HRG predictions of the $QC$ correlations normalised by $P_C$ contrasted with the PDG-HRG predictions [middle]. The fractional contribution of the $|Q|=2$ charm subsector to the total charm partial pressure as a function of temperature.  The combination $(\chi^{QC}_{22}-\chi^{QC}_{13})/2$ projects onto the $|Q|=2$ charm subsector for $T \leq T_{pc}$, and its ideal charm quark gas limit is represented by the red-solid line. See text for the definition of $x$ [right]. The open symbols
		represent the respective results from Ref. \cite{BAZAVOV2014210}. The yellow bands represent $T_{pc}$ with its uncertainty.}
	\label{fig:QC}
\end{figure}

The first-time calculation of the ${QC}$ correlations on the high-statistics datasets of the HotQCD Collaboration enables us to disentangle the contributions from different electrically-charged charm subsectors i.e., $|Q|=0,1 \text{ and }2$ in the hadronic phase in Fig.~\ref{fig:QC} [left]. Our lattice calculations of the fourth-order $QC$ correlations normalised by the total charm pressure show agreement with the QM-HRG model calculations, indicating the incomplete PDG records of the charmed hadrons carrying $|Q|=0,1 \text{ and } 2$. At a given temperature in Fig.~\ref{fig:QC} [middle], the ratio of QM-HRG prediction and PDG-HRG prediction increases with increasing $Q$-moments, implying that  $\chi^{QC}_{22}$ and $\chi^{QC}_{31}$ give evidence for the missing resonances. In particular, we see an enhancement over the PDG expectation in the fractional contribution of the $|Q|=2$ charm subsector to the total charm partial pressure for ${T<T_{pc}}$. In Fig.~\ref{fig:QC} [right], the observable $(\chi^{QC}_{22}-\chi^{QC}_{13})/2$ projects onto the $|Q|=2$ charm subsector for $T\leq T_{pc}$, and close to freeze-out, shows an enhancement, $x=1.8994$, over the PDG expectation in the fractional contribution of the $|Q|=2$ (or $\Sigma_c^{++}$) charm subsector to the total charm partial pressure. This enhancement is in agreement with the QM-HRG model calculations. 

\section{Absolute Charm Cumulants and the Continnum Limit}
\begin{minipage}[b]{0.45\textwidth}
	%\begin{figure}
	
	\includegraphics[width=\textwidth]{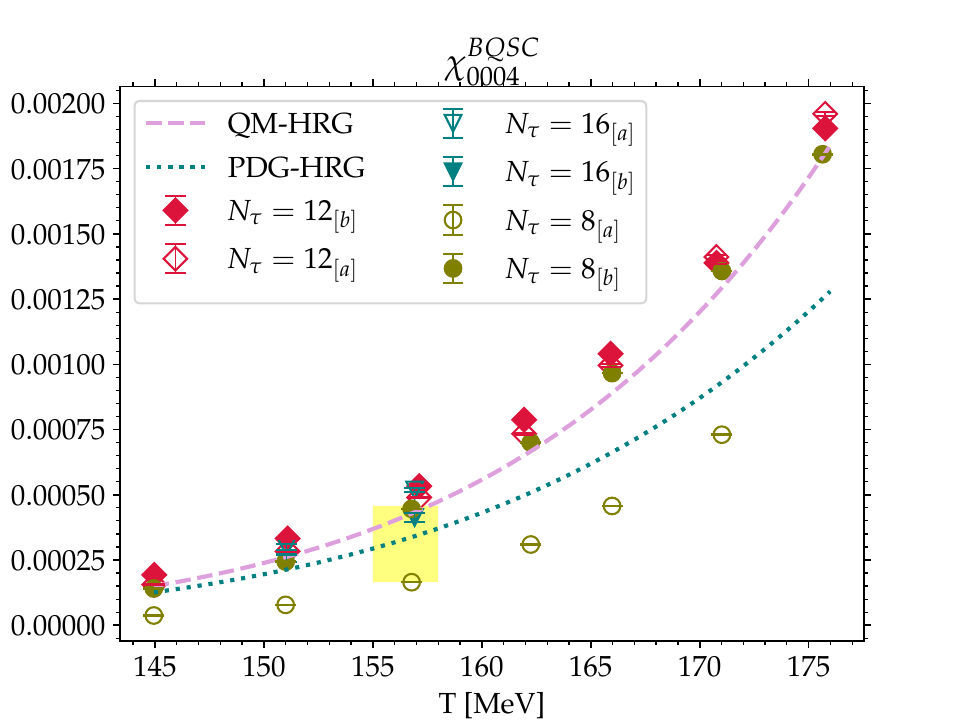}
	\captionof{figure}{The total charm pressure for different values of the temporal lattice extents calculated using two different LCP prescriptions. Also shown are the QM-HRG and the PDG-HRG predictions. The yellow band represents $T_{pc}$ with its uncertainty.}
	\label{fig:cont}
	%\end{figure}
\end{minipage}
\hspace{2 mm}
\begin{minipage}[b]{0.45\textwidth}
	As explained in Subsec. \ref{subsec:lcps}, to make a prediction on the basis of the absolute cumulants, we need to take the continuum limit. In this work, we approach the continuum limit in two different ways. The unfilled markers in Fig.~\ref{fig:cont} show $\chi^{C}_4$ results for LCP[a], whereas the filled ones represent $\chi^{C}_4$ results for LCP[b]. For $N_\tau>8$, results from two different LCPs converge and lie within $20\%$ of the QM-HRG prediction. It is worth pointing out that in contrast to the LCP[a] results for $N_\tau=8$, the LCP[b] results for $N_\tau=8$ lattices are already in the vicinity of the QM-HRG prediction. This is because $\chi^{C}_4$ is dependent upon the mass of the lightest hadron,
\end{minipage}

\noindent  which is D-meson. The lattice D-meson mass in turn depends upon the $am_c$ values shown  in Fig.~\ref{fig:para_comp}. The higher the $am_c$ value, the heavier the D meson, and thus smaller the total charm pressure. Therefore, the ordering of charm pressures based on different LCPs and $N_\tau$ values in Fig.~\ref{fig:cont} can be understood from the ordering of the $am_c$ values that went into the calculation of these charm pressures, and are shown in Fig.~\ref{fig:para_comp}: $\beta=[6.285-6.500]$ is relevant for $N_\tau=8$; $\beta=[6.712-6.910]$ is relevant for $N_\tau=12$; $\beta=[7.054-7.095]$ is relevant for $N_\tau=16$. 
\section{Conclusions and Outlook}
We show that the HRG description of the charm degrees of freedom breaks down at $T_{pc}$. Our results give evidence of deconfinement in terms of the appearance of charm quark-like excitations at $T_{pc}$. We show that the relevant charm degrees of freedom inside the QGP fall into three categories: meson-like, baryon-like and quark-like, and  the charm quarks become the dominant degree of freedom only at $T> 175$~MeV. Moreover, similar to ${T<T_{pc}}$ regime, our data suggests that for ${T>T_{pc}}$, the ${|Q|=2}$ charm subsector is solely composed of baryon-like states. Our results approach the ideal charm quark gas limit for $240 \text{ MeV}<T\leq340 \text{ MeV}$. Our findings show the existence of not-yet-discovered charmed hadronic states in all electrically-charged charm ($QC$) subsectors below $T_{pc}$. 

We find that the lattice cutoff effects in the absolute charm cumulants
are insignificant in the ratios of these absolute cumulants. These cutoff effects can be understood by considering different prescriptions for
fixing the lines of constant physics for the charm quark mass tunning. We find that the total charm pressure is dependent upon the bare charm quark mass value. Our future goal is to take the continuum limit of the absolute charm cumulants.
\section*{Acknowledgements}
This work was supported by The Deutsche Forschungsgemeinschaft (DFG, German Research Foundation) - Project number 315477589-TRR 211,
”Strong interaction matter under extreme conditions”.
The authors gratefully acknowledge the
computing time and support provided to them on the high-performance computer Noctua 2 at the NHR Center
PC2 under the project name: hpc-prf-cfpd. These are funded by the Federal Ministry of Education
and Research and the state governments participating on the basis of the resolutions of the GWK
for the national high-performance computing at universities (www.nhr-verein.de/unsere-partner).
Numerical calculations have also been performed on the
GPU-cluster at Bielefeld University, Germany. We thank the Bielefeld HPC.NRW team for their support. All the HRG calculations were performed using the AnalysisToolbox code developed by the HotQCD Collaboration \cite{Altenkort:2023xxi}.
\bibliographystyle{unsrt} 
\bibliography{merged}
%\begin{thebibliography}{99}
%\bibitem{...}
%....
%
%\end{thebibliography}

\end{document}